\documentclass[letter,traditabstract]{aa} 
\usepackage{graphicx}
\usepackage{natbib}
\usepackage{txfonts}
\begin{document}
   \title{Discovery of the correspondence between intra-cluster radio emission and a high pressure region detected through the Sunyaev-Zel'dovich effect}

   \author{C. Ferrari\inst{1}
   		 \and
          H.~T. Intema\inst{2}\thanks{Jansky Fellow of the National Radio Astronomy Observatory}
          \and
          E. Orr\`u\inst{3}
          \and
          F. Govoni\inst{4}
          \and
          M. Murgia\inst{4}
          \and
          B. Mason\inst{2}
          \and
          H. Bourdin\inst{5}
          \and
          K.~M. Asad\inst{6}
          \and
          P. Mazzotta\inst{5,7}
          \and
          M.~W. Wise\inst{8,9}
          \and
          T. Mroczkowski\inst{10,11}
          \and
          J.~H. Croston\inst{12}
          }

   \institute{Universit\'e de Nice Sophia Antipolis, CNRS, Observatoire de la C\^ote d'Azur, Laboratoire Cassiop\'ee, Nice, France, \email{chiara.ferrari@oca.eu}
         \and
         National Radio Astronomy Observatory, 520 Edgemont Road, Charlottesville, VA 22903-2475, USA
         \and
         Radboud University Nijmegen, Heijendaalseweg 135, 6525 AJ Nijmegen, The Netherlands
         \and
         INAF - Osservatorio Astronomico di Cagliari, Strada 54, Loc. Poggio dei Pini, 09012 Capoterra (Ca), Italy
         \and
      	Dipartimento di Fisica, Universit\`a degli Studi di Roma ``Tor Vergata'', via della Ricerca Scientifica 1, 00133 Roma, Italy
		 \and
		 Master student of the ``AstroMundus'' Erasmus Mundus Masters Course
		 \and
		 Harvard-Smithsonian Center for Astrophysics, 60 Garden St., Cambridge, MA 02138, USA
		 \and
		 ASTRON, PO Box 2, 7990 AA Dwingeloo, The Netherlands
		 \and
		 Astronomical Institute Anton Pannekoek, University of Amsterdam, Amsterdam, The Netherlands
		 \and
		 University of Pennsylvania, 209 S. 33rd St., Philadelphia, PA 19104, USA
		 \and
		 NASA Einstein Postdoctoral Fellow
		 \and
		 School of Physics and Astronomy, University of Southampton, Southampton, SO17 1BJ, UK}

   \date{Received 29 July 2011; accepted 6 October 2011}

 
  \abstract
   {We analyzed new 237~MHz and 614~MHz GMRT data of the most X-ray luminous galaxy cluster, RX\,J1347-1145. Our radio results are compared with the MUSTANG 90~GHz Sunyaev-Zel'dovich effect map and with re-processed {{\it Chandra}} and {{\it XMM-Newton}} archival data of this cluster. We point out for the first time in an unambiguous way the correspondence between a radio excess in a diffuse intra-cluster radio source and a hot region detected through both Sunyaev-Zel'dovich effect and X-ray observations. Our result indicates that electron re-acceleration in the excess emission of the radio mini-halo at the center of RX\,J1347-1145 is most likely related to a shock front propagating into the intra-cluster medium.} 
   {}

\authorrunning{Ferrari et al.}
\titlerunning{Discovery of radio emission in a shock heated region detected through SZE}
   \maketitle
%

\section{Introduction}

The existence of a non-thermal component (GeV electrons and $\mu$G magnetic fields) of the intra-cluster medium (ICM) has been revealed by the detection of diffuse radio sources that are not associated with active galaxies, but with the ICM. Through non-thermal studies of galaxy clusters we can estimate the cosmic-ray and magnetic field energy budget and pressure contribution to the ICM and also obtain clues about the cluster dynamical state and energy redistribution during merging events \citep[e.g.][]{2004JKAS...37..433S}. 

Up to now, only $\lesssim$ 10\% of known clusters are ``radio-loud'', i.e. show evidence of a diffuse non-thermal component in the radio band \citep[see][for recent reviews]{2009ASPC..407..223C,2011IAUS..274..340F}. Based on their physical properties, diffuse cluster radio sources are usually divided into three categories: halos, relics and mini-halos \citep[see, e.g.,][]{2008SSRv..134...93F}.  Radio halos are low surface brightness sources with a regular morphology that permeate the central cluster region and extend out to  $\gtrsim$ 1 Mpc. Relics have generally been found at the periphery of clusters and exhibit a wider range of morphologies. Mini-halos have sizes smaller than 500 kpc, and have been detected in the central regions of cool-core galaxy clusters, generally surrounding a powerful radio galaxy.

\begin{figure*}
   \centering
   \includegraphics[clip , width=0.31\textwidth , bb= 36 281 304 511]{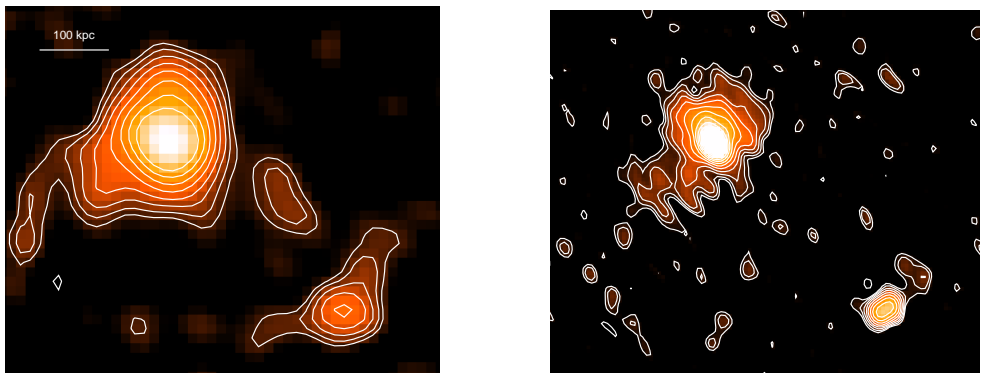}
   \hspace{.50in}
   \includegraphics[clip , width=0.31\textwidth , bb= 306 281 577 511]{Fig1.ps}
      \caption{GMRT radio images at 237~MHz (left -- resolution: $11.7 '' \times 9.3 ''$) and 614~MHz (right -- resolution: $4.8 '' \times 3.5 ''$) of RX\,J1347-1145 central region. The two images have the same physical scale. Total intensity radio contours, overlaid in white, are -2.7, 2.7, 3.8, 5.4, \dots mJy/beam at 237~MHz (rms noise level: 0.9 mJy/beam) and -0.3, 0.3, 0.42, 0.6, \dots mJy/beam at 614~MHz (rms noise level: 0.1 mJy/beam).
              }
         \label{fig:Radio}
   \end{figure*}

A common property of these three classes of objects is that the radiative lifetime of their relativistic electrons is much shorter than the timescale on which the radio-emitting plasma can fill the whole radio source volume \citep[e.g.][]{2001MNRAS.320..365B}. Different models have been proposed to explain the presence of cosmic-ray electrons in radio-loud clusters \citep[e.g.][]{2011MNRAS.412..817B,2011A&A...527A..99E}. Observational results are at present in favor of intra-cluster electron re-acceleration by shocks in the volume of radio relics, or turbulence in the case of halos and mini-halos \citep[see][and references therein]{2008SSRv..134...93F}. Most  Mpc-scale radio sources have been detected in luminous merging systems. Their radio power is generally correlated to the X-ray luminosity of the host cluster \citep[but see e.g.][and references therein for a few examples of outliers]{2011A&A...530L...5G}. The energy required to produce radio-emitting cosmic-rays comes therefore most likely from the huge gravitational energy released during cluster mergers ($\approx 10^{64}$ ergs). This is different for mini-halos, in which it has been suggested that a population of relic electrons ejected by the central AGN are most likely re-accelerated by MHD turbulence within the central cold cluster region \citep{2002A&A...386..456G}; this turbulence is possibly related to gas ``sloshing'' \citep[i.e. the oscillatory motion of the lowest entropy gas within the gravitational potential of merging clusters,][]{2008ApJ...675L...9M}. Unfortunately, our current observational knowledge of mini-halos is limited to only a handful of well-studied clusters \citep[see, e.g.,][and references therein]{2009A&A...499..679M}. More statistics as well as complementary detailed physical analyses of clusters hosting radio mini-halos are therefore required. 

We analyzed new GMRT observations of the most X-ray luminous cluster known -- RX\,J1347-1145 (hereafter RX\,J1347) -- that hosts a radio mini-halo \citep{1995A&A...299L...9S,2007A&A...470L..25G}. Our radio results are compared to millimeter and X-ray data. Particularly interesting for this work is the dynamical state of this cluster, for which a wealth of observational data exists at optical, X-ray, radio and mm wavelengths \citep[][and references therein]{2011arXiv1106.3489J}. Initially considered as the prototype of a relaxed cooling-flow cluster, RX\,J1347 has subsequently shown signatures of merging coming from millimeter observations of the Sunyaev-Zel'dovich effect \citep[SZE, e.g.][]{1999ApJ...519L.115P,2004PASJ...56...17K} and from higher sensitivity X-ray analyses \citep[e.g.][]{2002MNRAS.335..256A,2008A&A...491..363O}. The presence of a southeast (SE) substructure, characterized by a hot ICM (T $>$ 20 keV), has been pointed out and analyzed in detail through joint multi-wavelength and numerical studies \citep[][and references therein]{2011arXiv1106.3489J}. High resolution (10$''$) MUSTANG 90~GHz observations have recently confirmed the SZ signal at 20$''$ to the SE from the cluster center, indicating a strong, localized SZE decrement \citep{2010ApJ...716..739M,2011ApJ...734...10K} that is possibly associated to an ICM shock. 

The adopted cosmological parameters are $\Lambda$CDM (${\rm H}_0$=71 km ${\rm s}^{-1} {\rm Mpc}^{-1}$, $\Omega_{\rm m} = 0.27$, $\Omega_{\Lambda} = 0.73$). At the redshift of the cluster ($z$=0.451) 1$''$ corresponds to 5.74 kpc.


\section{Radio data reduction}

\begin{figure}[t]
\centering
\includegraphics[width=0.30\textwidth]{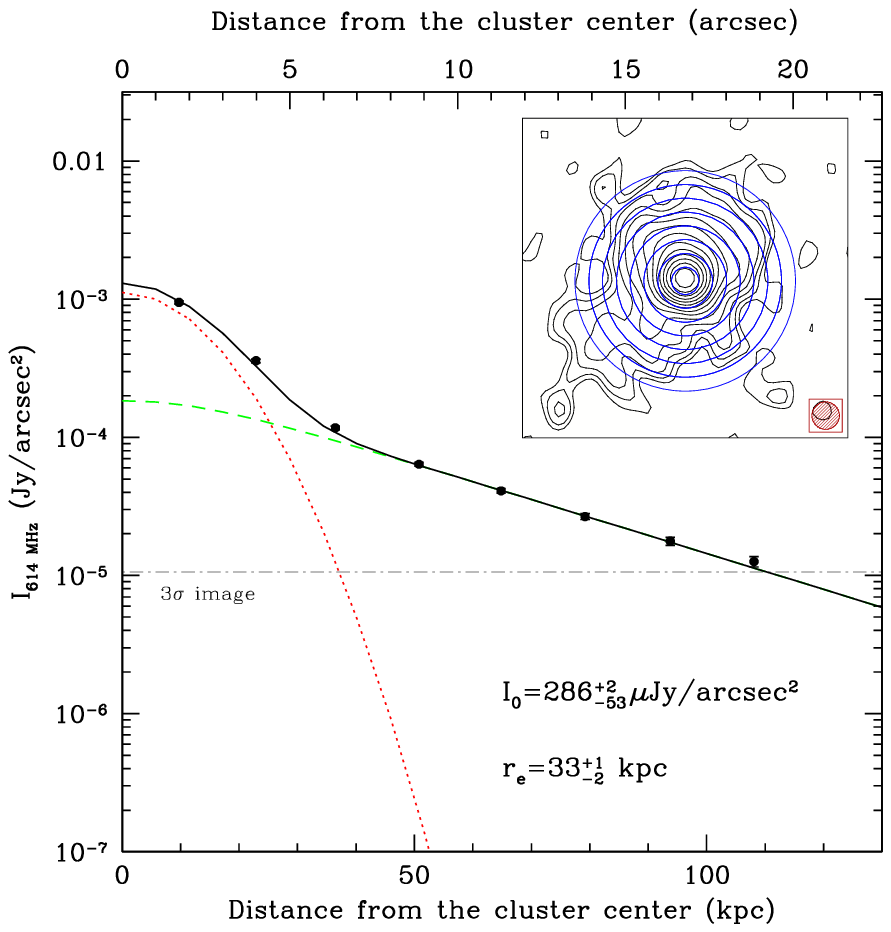}
\caption{Azimuthally averaged brightness profile of the radio emission in RXCJ1347 at 614~MHz. The profile is calculated in concentric annuli, as shown in the inset panel. The horizontal dashed-dotted line indicates the 3$\sigma_{\rm 614 MHz}$ noise level of the radio image. In our analysis we considered all data points above the $3\sigma_{\rm 614 MHz}$ noise level. The black line indicates the best-fit profile described by an exponential law (dashed line, Eq.\,\ref{eq:MH}) representing the mini-halo emission, and by a central Gaussian profile (dotted line, Eq.\,\ref{eq:PS}) representing the central point source.
}
\label{fig:profilo}
\end{figure}

GMRT observations of RX\,J1347 were obtained in the 240/610~MHz dual frequency mode. Visibilities were recorded every 16.8~seconds in 128~frequency channels covering 32~MHz of bandwidth at both frequencies. Data reduction was performed using the AIPS and SPAM software packages \citep{2009A&A...501.1185I}. After flagging, the remaining effective bandwidths are 6.25 and 13.5~MHz, centered on 237 and 614~MHz, respectively. The total effective time on-target is 12~hours. We used 3C\,147 as the primary flux and bandpass calibrator, adopting flux levels of 59.5 and 39.7 Jy at 237 and 614~MHz, respectively. The secondary calibrator 3C\,283 was used to determine slow-gain amplitude variations. The amplitude calibration results were applied to the target field data, followed by additional RFI flagging and frequency averaging to 25~channels of 0.25~MHz each at 237~MHz, and 18~channels of 0.75~MHz each at 614~MHz.

\begin{figure*}
   \centering
   \includegraphics[width=0.31\textwidth]{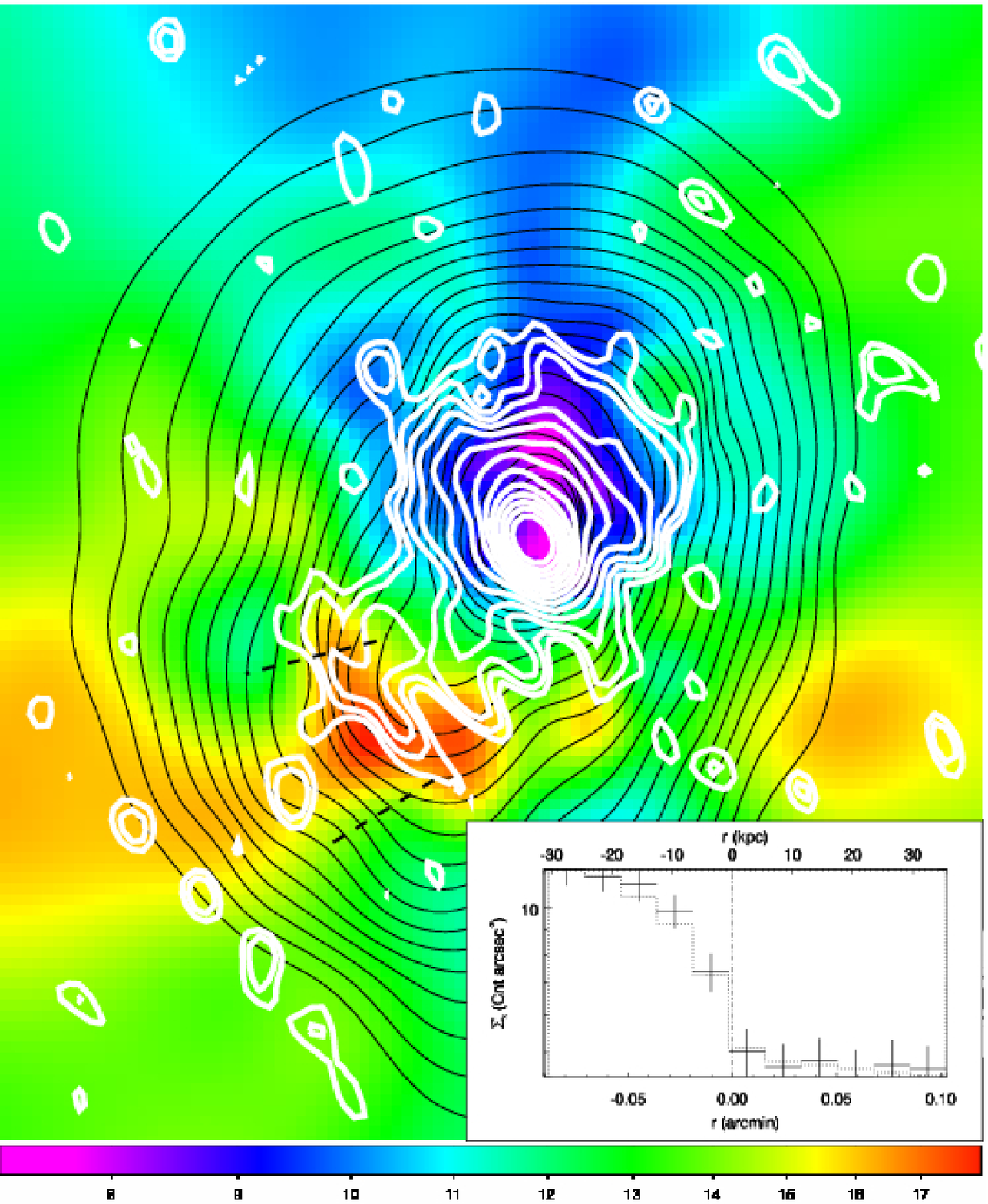}
   \hspace{.60in}
   \includegraphics[width=0.31\textwidth]{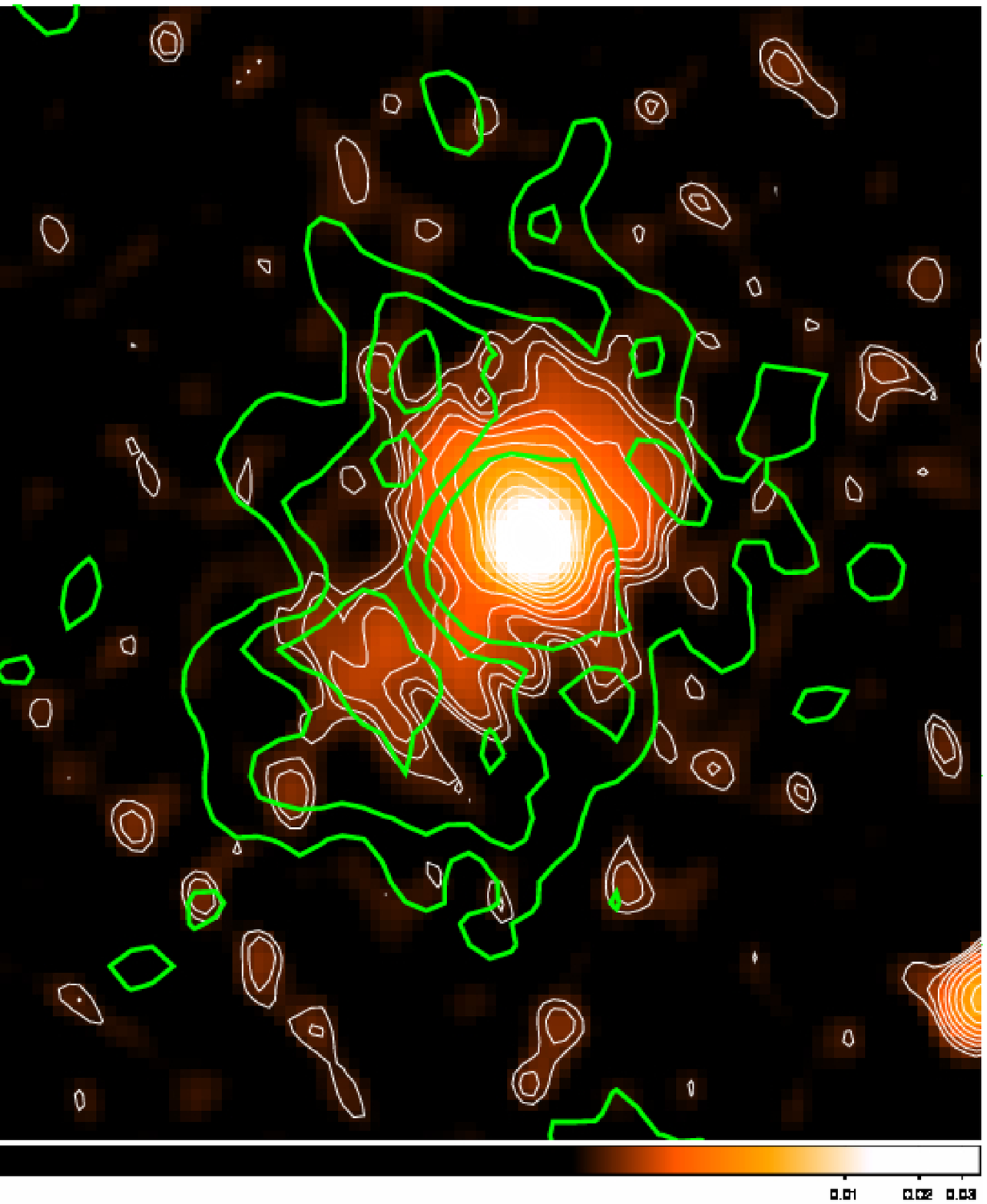}
      \caption{{\em Left:} {\it XMM-Newton} X-ray temperature map of RX\,J1347 in keV. X-ray iso-contours from the {\it Chandra} [0.5, 2.5] keV band image are superimposed in black. Total intensity radio contours are overlaid in white. They start at 3$\sigma$ level and are spaced by a factor of $\sqrt{2}$. The inset shows the ICM brightness profile corresponding to the sector illustrated by the dashed lines on the main image. A shock model corresponding to density and temperature jumps of $2.9^{+0.3}_{-0.2}$ and $0.5^{+0.8}_{-0.1}$ gives the fitted brightness profile indicated by dotted lines in the inset. {\em Right:} total intensity 614~MHz map and contours (white) of RX\,J1347. Contours of the MUSTANG SZE image of the cluster are overlaid in green \citep[levels: -1.5, -1.0 and -0.5 mJy/beam, from the inner contour outward, as in ][Fig. 6]{2010ApJ...716..739M}. The two panels have exactly the same physical scale.
              }
         \label{fig:multi-L}
   \end{figure*}

The target field data were phase-calibrated against a simple point source model derived from NVSS \citep{1998AJ....115.1693C} and WENSS \citep{1997A&AS..124..259R}, followed by several rounds of wide-field imaging, CLEAN deconvolution and self-calibration. Bright sources in the 614~MHz data were peeled to decrease the overall noise level. For the 237~MHz data we applied ionospheric calibration as implemented in SPAM. 

In Fig.\,\ref{fig:Radio} we show the final (uniform weighted) images at 237 and 614~MHz. Their respective synthesized beams and noise levels are  $11.7 '' \times 9.3 ''$ and $\sigma_{\rm 237~MHz}$ = 0.9 mJy/beam, and $4.8 '' \times 3.5 ''$ and $\sigma_{\rm 614~MHz}$ = 0.1 mJy/beam. 


\section{Results}

Radio emission at the center of RX\,J1347 results from a combination of a central point source and surrounding diffuse emission \citep{2007A&A...470L..25G}. In order to carefully separate the contribution of the mini-halo from that of the central radio galaxy and estimate the radio power of the diffuse source, we followed \citet{2009A&A...499..679M}. The total brightness profile of the radio emission at the center of the cluster was fitted taking into account a central point source ($I_{\rm PS}$) plus the radio mini-halo diffuse emission ($I_{\rm MH}$):

\begin{equation}
I(r)=I_{\rm PS}(r)+I_{\rm MH}(r). 
\end{equation} 	

The profile of the point and diffuse sources were adopted to be a Gaussian and an exponential law, respectively:

\begin{equation}
I_{\rm PS}(r)=I_{0_{\rm PS}}~{\rm e}^{-(r^2/2\sigma_{\rm PS}^2)} \label{eq:PS},
\end{equation} 	

\begin{equation}
I_{\rm MH}(r)=I_{0_{\rm MH}}~{\rm e}^{-r/r_{e}} \label{eq:MH}. 
\end{equation}

In Fig.\,\ref{fig:profilo} we show the azimuthally averaged radio brightness profile at 614 MHz (i.e. the higher resolution of the two GMRT maps) traced down to a level of 3$\sigma_{\rm 614~MHz}$. The radio image was convolved to 5$''$ resolution. The annuli, as shown in the inset panel, are as wide as the half FWHM beam. The S/N ratio of this map is sufficient to allow a very good separation between the point source and diffuse emission. The best-fit model is shown as a continuous black line in the right panels of Fig.\,\ref{fig:profilo}. The mini-halo contribution is indicated by the dashed line. Overall the mini-halo clearly extends from the central point source and it is well fitted by the exponential model. The best fit of the exponential model yields a central brightness of $I_{\rm 0}$=$286_{-53}^{+2}$ $\mu$Jy/arcsec$^{2}$ and $r_{\rm e}$=$33_{-2}^{+1}$ kpc. 
   
The flux density of the mini-halo at 614 MHz integrated up to 3 $r_{\rm e}$ is $S_{\rm 614 MHz}$=$48 \pm 2$ mJy, while the flux density calculated up to the size of the diffuse brightness emission containing the 3$\sigma_{\rm 614 MHz}$ radio isophotes is $S_{\rm 614 MHz}$=$50 \pm 2$ mJy. We estimated the flux density at 237 MHz up to 3$\sigma_{\rm 237 MHz}$ level from the map, resulting in $S_{\rm 237 MHz}$=$131 \pm 6$ mJy. Following \citet{2011arXiv1106.6228V}, we also subtracted the central point source in frequency space by obtaining flux measures that agree very well with those derived through the fitting procedure. We then estimated the 237 and 614 MHz fluxes of the diffuse source from the point source subtracted maps within the 614~MHz 3$\sigma$ contours and derived a mean spectral index of $\alpha_{237}^{614} \simeq 0.98 \pm 0.05$\footnote{$S_{\nu} \propto \nu^{-\alpha}$} for the mini-halo. The central point source has a flux of $55 \pm 4$ mJy at 237~MHz and $32 \pm 2$ mJy at 614~MHz.

We compared our radio observations to cluster gas brightness and temperature maps obtained from archival X-ray observations through B2-spline wavelet imaging and spectral imaging analyses, as detailed in \citet{2008A&A...479..307B}. We used the Chandra observation to map the gas brightness at a 1$''$ angular resolution. We took advantage of the larger effective area of {\it XMM-Newton} at high energy to map the gas temperature from 3 $\sigma$ thresholding of the wavelet coefficients, investigating significant features within a resolution range of 4 to 32 arcsec.

An elongation in the radio mini-halo morphology is evident both in the 614~MHz and 237~MHz GMRT maps at more than 5$\sigma$ level (see Fig.\,\ref{fig:Radio}). It lies in the SE X-ray substructure, which is evident in {\it Chandra} data (black contours in the left panel of Fig.\,\ref{fig:multi-L}). The radio excess, particularly clear and resolved for the first time in the 614~MHz map, corresponds exactly to the position of the hottest region SE of the cluster core detected by MUSTANG SZE imaging (inner green contour in the right panel of Fig.\,\ref{fig:multi-L}, which indicates the strongest SZ decrement). In the same region, our {\it XMM-Newton} temperature map reveals a hot (T $\gtrsim$ 17 keV) structure, delimited to the SE by a surface brightness edge that might be a shock front (see left panel of Fig.\,\ref{fig:multi-L}). These results agree with previous X-ray analyses \citep{2002MNRAS.335..256A} in concluding that the strong SZ decrement region of RX\,J1347 is presumably shock-heated. Aside from the SE excess, the rest of the mini-halo radio emission seems to be well confined within the cold central part of the cluster, surrounded by the high-pressure gas pointed out by SZE observations, and shown in the right panel of Fig.\,\ref{fig:multi-L} by the two external contours \citep[see also Fig. 6 in ][]{2010ApJ...716..739M}.

    \begin{figure}
   \centering
   \includegraphics[width=0.31\textwidth]{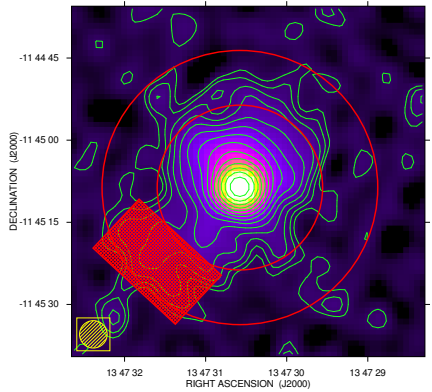}
      \caption{614~MHz map of RX\,J1347 convolved to 5$''$ resolution and total intensity contours starting from 3 $\sigma$ level. The mean surface brightness of the annulus indicated in this image was estimated excluding (or including only) the excess region within the rectangular region (see text).}
         \label{fig:annulus}
   \end{figure}

To obtain a rough estimate of the radio flux in the SE excess region of the mini-halo, we considered an annulus containing the radio elongation (see Fig.\,\ref{fig:annulus}). We then estimated the mean surface brightness in the higher-resolution 614~MHz map a) within the annulus, excluding the excess region ($<I> = 0.25 \pm 0.02$ mJy/beam over $\sim$37.1 beam area), and b) only within the excess region (rectangular area in Fig.\,\ref{fig:annulus}, $<I> = 0.63 \pm 0.03$ mJy/beam over $\sim$8.8 beam area). The net excess in the radio surface brightness of the SE cluster region is therefore $<I> = 0.38 \pm 0.04$ mJy/beam, corresponding to a radio flux of $3.3 \pm 0.3$ mJy at 614~MHz.

\section{Conclusions}

We showed for the first time a clear correspondence between an excess emission in the radio mini-halo at the center of RX\,J1347 and a high pressure ICM region revealed by SZE observations at a similarly high angular resolution \citep[$\lesssim$ 10$''$,][]{2010ApJ...716..739M}. Possible evidence of radio emission excess in the SE direction were pointed out by higher radio frequency observations \citep{2007A&A...470L..25G}, but without the sensitivity and resolution offered by GMRT data. Our radio observations uniquely allow us to probe the exact spatial coincidence with the SZ decrement detected at 90 GHz by \citet{2010ApJ...716..739M} and with the hot ICM region shown in our temperature map. If we exclude the SE elongation in the mini-halo morphology, the rest of the diffuse radio source is confined within the colder central region. 

Electron re-acceleration in the mini-halo at the center of RX\,J1347 can be related to turbulence produced by gas sloshing that is typical of disturbed clusters \citep{2008ApJ...675L...9M}. However, additional physical mechanisms are needed to explain the detected radio excess. The possible SE shock -- most likely confirmed by our X-ray analysis -- could be responsible for local intra-cluster electron re-acceleration \citep{1998A&A...332..395E}. An alternative hypothesis is that the hot gas in the SE of the cluster is related to adiabatic gas compression, which amplifies the intra-cluster magnetic field intensity and increases the energy of radio emitting electrons, resulting in a higher synchrotron emissivity \citep[e.g.][]{2001A&A...366...26E}. We estimate that in this second case, a 15\% volume compression is required to justify the observed surface brightness excess.

The diffuse radio source at the center of RX\,J1347 presents intermediate properties between ``classical'' radio mini-halos and relics, cosmic-ray acceleration in this system resulting from the combination of different physical mechanisms. Other mini-halos could present similar properties when analyzed through such detailed multi-wavelength observations. The implications of this study indeed go beyond the single case of RX\,J1347, because we clearly show the perspectives opened by new high-resolution multi-wavelength observations for cluster studies. In particular our study highlights the importance of combining good resolution ($\sim$ 5$''$-10$''$) observations at a) low radio frequencies ($<$ 1 GHz), which are best suited for the detection of diffuse intra-cluster radio sources that are characterized by steep synchrotron spectra, and b) millimeter observations, which are favored for depicting shock regions, given the linear dependence to the ICM pressure.  

In the next few years joint higher-sensitivity and/or resolution observations derived from X-ray, mm and radio facilities (e.g. NuSTAR, MUSTANG--2, Planck, LOFAR, EVLA, \dots) will allow us to surpass phenomenological comparisons of thermal and non-thermal cluster properties, and to achieve a clearer characterization of the physics driving non-thermal intra-cluster phenomena. Most likely, traditional classifications of diffuse intra-cluster radio sources will lead to a more general view of multi-scale, complex radio emission, which is difficult to classify in a general scheme and deeply connected to the dynamical history and thermal evolution of each cluster.

\begin{acknowledgements}
We thank the anonymous referee for her/his useful comments. We are grateful to Giulia Macario for helpful suggestions in the radio data reduction phase. We warmly thank Monique Arnaud for very useful discussions. We would like to thank the staff of the GMRT that made these observations possible. GMRT is run by the National Centre for Radio Astrophysics of the Tata Institute of Fundamental Research. CF acknowledges financial support by the ``{\it Agence Nationale de la Recherche}'' through grant ANR-09-JCJC-0001-01.  This work was supported by the BQR program of {\it Observatoire de la C\^ote dÕAzur} and by {\it Laboratoire Cassiop\'ee} (UMR\,6202).
\end{acknowledgements}

\bibliographystyle{aa}
\bibliography{RXJ1347}

\end{document}